\documentstyle[preprint,aps]{revtex} 
\begin{document} 

\draft 

\preprint{UTPT-96-10} 

\title{A Solution to the Cosmological Constant Problem}

\author{J. W. Moffat} 

\address{Department of Physics, University of Toronto,
Toronto, Ontario, Canada M5S 1A7} 

\date{\today}

\maketitle 

\begin{abstract}%
The fluctuations of the vacuum energy are treated as a non-equilibrium process
and a stochastic model for the cosmological constant is presented, which
yields a natural explanation for the smallness or zero value of the constant
in the present epoch and its large value in an era of inflation in the early
universe.
\end{abstract} 

\pacs{ }

It is widely considered that the problem of the cosmological constant
has not been satisfactorily explained. The problem is this:
the vacuum fluctuations of particle fields add up to a contribution which is
much larger than the observational limits by many orders of magnitude. Attempts
using quantum cosmology, based on e.g., Euclidean path integrals and 
wormholes, have been criticized for various technical reasons, which question it as a
possible solution to the problem\cite{Carroll}.
Unbroken supersymmetry can ``protect" the constant $\Lambda$ from becoming
non-zero, but supersymmetry is badly broken in the
real universe at an energy $>1$ TeV. No other symmetry is known to exist in
Nature that guarantees that $\Lambda$ remains zero.
Recently, a solution using the Jordan, Brans-Dicke theory of gravitation
was proposed in which the cosmological ``constant" is a smooth function
of time\cite{Moffat}. 

It is believed by many that the problem would be solved within a physically
consistent theory of quantum gravity. Since such a theory does not presently
exist, we shall propose a possible solution to the problem based on treating
the vacuum energy as a stochastic system. We describe this system by a
phenomenological equation for the cosmological constant $\Lambda$, which
appears in Einstein's field equations:
\begin{equation}
R_{\mu\nu}-\frac{1}{2}g_{\mu\nu}R+\Lambda g_{\mu\nu}=8\pi GT_{\mu\nu}.
\end{equation}
The cosmological constant enters through the vacuum energy density:
\begin{equation}
T_{V\mu\nu}=-\rho_Vg_{\mu\nu}=-\frac{\Lambda_V}{8\pi G}g_{\mu\nu}.
\end{equation}
Today, $\Lambda$ has the incredibly small value, $\Lambda 
< 10^{-46}\,\,\hbox{GeV}^4$,
whereas generic inflation models require that $\Lambda$ has a large value
during the inflationary epoch. This is the source of the cosmological constant
problem.

Let us treat the vacuum energy as a fluctuating environment and 
consider $\Lambda$ as a variable characterizing the state of this system,
which is determined by the equation\cite{Verhulst,Horsthemke,Horsthemke2}:
\begin{equation}
\label{Lambdaequation}
{\dot \Lambda}=\alpha\Lambda-\Lambda^2,
\end{equation}
where ${\dot \Lambda}=d\Lambda/dt$, $\alpha$ is a parameter which corresponds
to the difference between
the rate of creation and annihilation of particles in the vacuum. The
second term
is a self-restriction term which limits the growth of $\Lambda$.  The solution
of (\ref{Lambdaequation}) is given by
\begin{equation}
\Lambda(t)=\Lambda(0)\exp(\alpha t)\{1+\Lambda(0)
[(\exp(\alpha t)-1)/\alpha]\}^{-1}.
\end{equation}
At $\alpha=0$, there is only one stable stationary state solution ${\bar \Lambda}=0$,
and at this point the solution becomes unstable and bifurcates into a new
branch of steady stable solutions, ${\bar \Lambda}=\alpha$. At $\alpha=0$ the
system undergoes a second-order phase transition.

Let us consider the situation in which the vacuum fluctuations are rapid compared
with $\tau_{\hbox{macro}}=\alpha^{-1}$, which defines the macroscopic scale of
time evolution. We shall assume that the parameter $\alpha$ can be written as
$\alpha_t=\alpha+\sigma\xi_t$, in which $\alpha$ is the average value,
$\xi_t$ is Gaussian noise and $\sigma$ measures the intensity of the vacuum
fluctuations. Then, we write Eq.(\ref{Lambdaequation}) as
\begin{equation}
d\Lambda_t=(\alpha\Lambda_t-\Lambda_t^2)dt+\sigma\Lambda_t dW_t
=f(\Lambda_t)+\sigma g(\Lambda_t)dW_t,
\end{equation}
where $dW_t$ is a Wiener process. We shall use the Ito integral to describe the
diffusion process, although the Stratonovich integral would predict the same
qualitative results\cite{Gihman}. The probability density $p(x,t)$ satisfies
the Fokker-Planck equation:
\begin{equation}
\label{Fokker}
\partial_t p(x,t)=-\partial_x[(\alpha x-x^2)p(x,t)]+\frac{\sigma^2}{2}
\partial_{xx}(x^2 p(x,t)).
\end{equation}
The diffusion process is restricted to the positive real half line and $0$ and
$\infty$ are intrinsic boundaries, because $g(0)=0$ and $f(\infty)=-\infty$. The
probability
of the diffusion process reaching infinity as $t\rightarrow\infty$ is zero, since
infinity is a natural boundary. Moreover, zero is a natural boundary if $\alpha >
\sigma^2/2$\cite{Horsthemke2}, so neither boundary is accessible and no boundary
conditions need be imposed on the Fokker-Planck equation. For $\alpha <
\sigma^2/2$, it can be shown that zero is an {\it attracting} boundary.

The stationary-state solution for the probability density, $p_s(x)$, of 
Eq.(\ref{Fokker}) is given by
\begin{equation}
p_s(x)=Nx^{(2\alpha/\sigma^2)-2}\exp\biggl(-\frac{2x}{\sigma^2}\biggr).
\end{equation}
The normalization constant $N$ is
\begin{equation}
N^{-1}=\biggl[\biggl(\frac{2}{\sigma^2}\biggr)^{2(\alpha/\sigma^2)-1}\biggr]^{-1}
\Gamma\biggl(\frac{2\alpha}{\sigma^2}-1\biggr).
\end{equation}
If $p(x,t)$ is integrable between $0$ and $\infty$, then a stationary state solution
exists when $\alpha> \sigma^2/2$. If it does not exist, then the probability density
will be concentrated at zero, i.e., $p_s(\Lambda) = \delta(\Lambda)$ for $\alpha <
\sigma^2/2$.

The extrema of $p_s$ play the role of order parameters for non-equilibrium
phase transitions, and they are determined by the equation:
\begin{equation}
\Lambda_m^2-(\alpha -\sigma^2)\Lambda_m=0.
\end{equation}
They are given by $\Lambda_{m1}=0$ and $\Lambda_{m2}=\alpha-\sigma^2$
($\alpha > \sigma^2$). The maximum $\Lambda_{m2}$ always exists, while
the maximum $\Lambda_{m1}$ exists for $0<\alpha<\sigma^2$. Two transition
points exist: one at $\alpha=\sigma^2/2$ and one at $\alpha=\sigma^2$. We
therefore have the following
situation: (a) If $\alpha < \sigma^2/2$, then zero is a stable stationary point
for $\Lambda$. (b) The point $\alpha =\sigma^2/2$ is a transition point, since
$\Lambda=0$ becomes unstable and a new stationary probability density is
produced. (c) The stationary density becomes divergent at $\Lambda=0$ when
$\sigma^2/2< \alpha < \sigma^2$. Although zero is no longer a stable stationary
point, it remains the most probable value. (d) For $\alpha =\sigma^2$, the
probability density $p_s(\Lambda)$ undergoes a transition, in which $\Lambda$ can
take on large values by increasing the intensity of the fluctuations, even though
the average state of the vacuum is kept constant.

For $0 < \alpha < \sigma^2/2$, the vacuum fluctuations dominate over the growth
or decline of $\Lambda$, although the value zero is still the most probable value
for $\Lambda$, since the distribution function has a vertical slope at
$\Lambda=0$. Because we are using a continuous variable, $\Lambda$ never
reaches the boundary zero in a finite time. This model breaks down when
$\Lambda$ is vanishingly small and the probability of having $\Lambda=0$ is
defined for $0<\epsilon<<1$ and $0<\alpha<\sigma^2/2$ by\cite{Horsthemke2}:
\begin{equation}
\hbox{lim}_{t\rightarrow\infty}\int^\epsilon_0\rho(x,t)dx.
\end{equation}

When $\alpha > \sigma^2/2$, the growth of
$\Lambda$ dominates the influence of the vacuum fluctuations, and in the
neighborhood of zero the probability of $\Lambda=0$ drops to zero. 
At $\alpha=\sigma^2/2$ real growth of $\Lambda$ becomes possible corresponding
to a change from a degenerate random variable for steady-state behavior to
a stochastic variable; the boundary at $\alpha=0$ switches from attracting to
natural. The transition point $\alpha=\sigma^2$ 
corresponds to a qualitative change in the stochastic variable $\Lambda$
with no change in the nature of the boundary. The probability of $\Lambda=0$
drops abruptly to zero. This phenomenon is inherently nonlinear.

We now see the following scenario emerging from our model. In the inflation
era, the intensity of vacuum fluctuations is large and $\alpha > \sigma^2$,
causing a second-order phase transition and a maximum in $\Lambda$ not
near zero. This corresponds to the large vacuum energy needed to drive
inflation\cite{Linde}. As the universe expands the intensity of vacuum fluctuations
decreases
and for $0<\alpha<\sigma^2/2$ or $\sigma^2/2<\alpha<\sigma^2$ the probability
density is largest when $\Lambda$ is non-vanishing and small, which can lead to a
current value of
$\Lambda_0$ that can be used to fit the observational data. 
If the stationary probability density $p_s$ does not exist for $\alpha<\sigma^2/2$,
then $\Lambda=0$
is a stationary point; the drift and diffusion vanish simultaneously for $\Lambda=0$
and $p_s(\Lambda)=\delta(\Lambda)$. This corresponds to the case when 
$\Lambda$ is vanishingly small.

Thus, our model provides a natural explanation, in terms of non-equilibrium
stochastic processes in an expanding universe, for the behavior of $\Lambda$ needed
to fit observational data and still be consistent with inflationary models.

According to general relativity, the equation that governs the expansion factor
$R(t)$ is given by
\begin{equation}
H^2\equiv\biggl(\frac{{\dot R}}{R}\biggr)=\frac{8\pi G}{3}\rho_M
+\frac{\Lambda}{3}-\frac{k}{R^2},
\end{equation}
where $\rho_M$ is the mass density; $k=-1,0,+1$ and $H$ is the Hubble constant,
whose observable value at present time $t_0$ is denoted by $H_0$. We define
\begin{equation}
\Omega_{\hbox{tot}}\equiv\Omega_M+\Omega_{\Lambda}=1-\Omega_k.
\end{equation}

Many observational tests can constrain $\Omega_{\Lambda}$,
but the gravitational lensing method provides the most direct constraint with the
result: $\Omega_{0,\Lambda} <0.75$\cite{Turner}. A recent analysis of the
cosmological data showed that for $\Omega_{\Lambda}=0.65\pm 0.1,
\Omega_M=1-\Omega_{\Lambda}$ and a small tilt: $0.8<n<1.2$, models exist
which are consistent with the available data and an inflationary spatially flat
universe\cite{Steinhardt}.

\acknowledgments

This work was supported by the Natural Sciences and Engineering Research
Council of Canada.

\end{document}